\documentclass[american,aps,pra,reprint, superscriptaddress, onecolumn]{revtex4-1}
\usepackage{amsmath,amscd,amsthm,amssymb}
\usepackage{mathrsfs}
\usepackage{graphicx}
\usepackage{epsf,epstopdf}
\usepackage[all]{xy}
\usepackage[unicode=true,pdfusetitle, bookmarks=true,bookmarksnumbered=false,bookmarksopen=false, breaklinks=false,pdfborder={0 0 0},backref=false,colorlinks=false] {hyperref}
\hypersetup{colorlinks=true,linkcolor=myurlcolor,citecolor=myurlcolor,urlcolor=myurlcolor}
\usepackage{graphics,epstopdf,graphicx, amsthm, amsmath, amssymb, times, braket, color, bm}
\usepackage{caption3}
\usepackage{float}
\usepackage[caption = false]{subfig}
\usepackage{cleveref}
\usepackage{makecell}

\definecolor{myurlcolor}{rgb}{0,0,0.7}

\theoremstyle{plain}

\providecommand{\theoremname}{Theorem}
\newcommand*{\myproofname}{Proof}

\makeatother

\theoremstyle{definition}

\theoremstyle{remark}

\usepackage{amscd}

\newcommand{\beq}{\begin{equation}}
\newcommand{\eeq}{\end{equation}}
\newcommand{\ba}{\begin{array}}
\newcommand{\ea}{\end{array}}
\newcommand{\bea}{\begin{eqnarray}}
\newcommand{\eea}{\end{eqnarray}}

\begin{document}

\title{Remarks on controlled measurement and
quantum algorithm for calculating Hermitian conjugate}

\author{Edward B. Fel'dman}
\email{efeldman@icp.ac.ru}
\affiliation{Federal Research Center of Problems of Chemical Physics and Medicinal Chemistry RAS,
Chernogolovka, Moscow reg., 142432, Russia}

\author{Alexander I. Zenchuk}
\email{zenchuk@itp.ac.ru}
\affiliation{Federal Research Center of Problems of Chemical Physics and Medicinal Chemistry RAS,
Chernogolovka, Moscow reg., 142432, Russia}

\author{Wentao Qi}
\email{qiwt5@mail2.sysu.edu.cn}
\affiliation{Institute of Quantum Computing and Computer Theory, School of Computer Science and Engineering, Sun Yat-sen University, Guangzhou 510006, China}

\author{Junde Wu}
\email{wjd@zju.edu.cn}
\affiliation{School of Mathematical Sciences, Zhejiang University, Hangzhou 310027, PR~China}

\begin{abstract}

We present two new aspects for the recently proposed algorithms for matrix manipulating based on the special encoding the matrix elements into the superposition state of a quantum system. First aspect is the controlled measurement which allows to avoid the problem of small access probability to the required ancilla state at the final step of algorithms needed to remove the garbage of the states. Application of controlled measurement to the earlier developed algorithm is demonstrated. The second aspect is the algorithm for calculating the Hermitian conjugate of an arbitrary matrix, which supplements the algorithms proposed earlier. The appropriate circuits are presented.

{\bf Keywords:} quantum algorithm, quantum circuit, determinant and inverse of matrix, linear algebraic systems, depth of algorithm, quantum measurement
\end{abstract}
\maketitle

\section{Introduction}

Quantum algorithms for matrix manipulations represent a large family of quantum algorithms that pretend to provide advantages in comparison with their classical counterparts. To the first algorithms of this type we can refer the
HHL algorithm for solving systems of linear algebraic equations \cite{HHL,HHL1,HHL2,HHL3,HHL4,HHL5,HHL6,HHL7},
in which the central  point is the conditional rotation of the state of the one-qubit ancilla (that perhaps requires appealing to the classical tool)  followed by  measurement of ancilla state with output 1.
There is another  family  of algorithms for matrix manipulations based on the Trotterization method and the Baker-Champbell-Hausdorff formula for exponentiating matrices \cite{ZhaoL1}.

Recently we propose  alternative algorithms for matrix manipulations, such as inner product, sum and product of two  matrices, calculation of determinant, matrix inversion, and solving linear systems, based on special unitary transformations applied to the quantum system   \cite{QZKW_arxive2022}.
In those algorithms we use the encoding of matrix elements into the pure superposition states of certain subsystems. The realization of those unitary operators in terms of the basic quantum operations was explored later
 in \cite{ZQKW_arxiv2023} (inner product, matrix addition and multiplication) and then in \cite{ZBQKW_2024}
 (calculating determinant and inverse matrix and linear system solver). Those unitary operators are based on the  multi-qubit controlled operators and  include the ancilla measurement for garbage removal.

The principal disadvantage of the algorithms proposed in \cite{ZQKW_arxiv2023,ZBQKW_2024} is the small probability of obtaining the required ancilla state after measurement. This probability increases exponentially with the dimension of the considered matrix.  In our paper, we propose the so-called controlled measurement
which removes the above disadvantage. In addition, we propose the algorithm for calculating the Hermitian conjugate of an arbitrary matrix.

The paper is organized as follows. The controlled measurement with application to the known algorithms is proposed in Sec.\ref{Section:CM}.
The quantum algorithm for calculating the Hermitian conjugate of an arbitrary matrix is considered in Sec.\ref{Section:HC}. Conclusions are presented in Sec. \ref{Section:conclusions}.

\section{Controlled measurement}
\label{Section:CM}
We consider  the $n^{(S)}$ qubit subsystem $S$ and $n^{(R)}$ qubit subsystem $R$. 
The state of the whole system $S\cup R$ reads
\begin{eqnarray}
|\chi_0\rangle =\sum_{i=0}^{N^{(S)}-1} \sum_{j=0}^{N^{(R)}-1} a_{ij}|i\rangle_S |j\rangle_R ,\;\;\sum_{i=0}^{N^{(S)}-1} \sum_{j=0}^{N^{(R)}-1} |a_{ij}|^2=1.
\end{eqnarray}
Our purpose is to construct the state
\begin {eqnarray}
\sim \sum_{i=0}^{N^{(S)}-1} \left(\sum_{j=0}^{N^{(R)}-1}a_{ij}\right) |i\rangle_S,
\end{eqnarray}
i.e., realize the (normalized) sum of $a_{ij}$ over the second subscript.
Applying the Hadamard transformation $H$ to each qubit of $R$ and collecting all $H$   in the operator $H_R$ we obtain
\begin{eqnarray}
|\chi_1\rangle=H_R |\chi_0\rangle =\frac{1}{2^{n^{(R)}/2}} \sum_{i=0}^{N^{(S)}-1} \sum_{j=0}^{N^{(R)}-1} a_{ij}|i\rangle_S |0\rangle_R
  + |garbage\rangle_{SR}.
\end{eqnarray}
Now, to label the garbage, we introduce the one-qubit ancilla $A_1$ in the ground state and apply the controlled operator
\begin{eqnarray}\label{WRA}
W^{(1)}_{RA_1}=|0\rangle_R \, _R\langle 0| \otimes \sigma^{(x)}_{A_2} + (I_R -  |0\rangle_R \, _R\langle 0| ) \otimes I_{A_1}
\end{eqnarray}
to the state
$|\chi_1\rangle |0\rangle_{A_1} $:
\begin{eqnarray}\label{chi2}
|\chi_2\rangle=W^{(1)}_{RA_1}|\chi_1\rangle |0\rangle_{A_1} =\frac{1}{2^{n^{(R)}/2}} \sum_{i=0}^{N^{(S)}-1} \sum_{j=0}^{N^{(R)}-1} a_{ij}|i\rangle_S |0\rangle_R |1\rangle_{A_1} + |garbage\rangle_{SR}|0\rangle_{A_1}.
\end{eqnarray}
Hereafter $\sigma^{(x)}_P$ means the $\sigma^{(x)}$ Pauli operator applied to the one-qubit subsystem $P$.

\begin{figure}[ht]
    \includegraphics[width=0.5\textwidth]{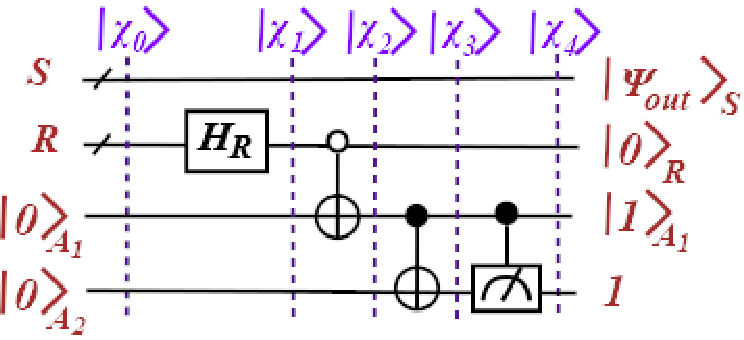}
    \caption{Application of controlled measurement.}
\label{Fig:CM}
\end{figure}

If we need to remove the garbage via ancilla measurement than we face the problem of small probability for obtaining 1 in result of measurement. This probability $\sim \frac{1}{2^{n^{(R)}}}$ and decreases exponentially with $n^{(R)}$.
To avoid this effect we introduce the controlled measurement of the ancilla state. To this end we introduce one more one-qubit ancilla $A_2$ in the ground state and the following
C-NOT:
\begin{eqnarray}\label{cnot}
W^{(2)}_{A_1A_2}=|1\rangle_{A_1} \, _{A_1}\langle 1| \otimes \sigma^{(x)}_{A_2} + |0\rangle_{A_1} \, _{A_1}\langle 0| \otimes I_{A_2} .
\end{eqnarray}
Applying  $W^{(2)}_{A_1A_2}$ to $|\chi_2\rangle |0\rangle_{A_2}$ we obtain
\begin{eqnarray}
|\chi_3\rangle=W^{(2)}_{A_1A_2}|\chi_2\rangle |0\rangle_{A_2} =\frac{1}{2^{n^{(R)}/2}} \sum_{i=0}^{N^{(S)}-1} \sum_{j=0}^{N^{(R)}-1} a_{ij}|i\rangle_S |0\rangle_R
|1\rangle_{A_1}|1\rangle_{A_2} + |garbage\rangle_{SR}|0\rangle_{A_1}|0\rangle_{A_2}.
\end{eqnarray}
Now we can introduce the following controlled measurement:
\begin{eqnarray}\label{CM}
M_{A_1A_2} = |1\rangle_{A_1} \, _{A_1}\langle 1| \otimes M_{A_2} + |0\rangle_{A_1} \, _{A_1}\langle 0|\otimes I_{A_2},
\end{eqnarray}
where $M_{A_2}$ is the measure operator applied to the ancilla $A_2$. Notice that the operator $M_{A_1A_2}$ is not unitary one due to the measurement $M_{A_2}$.
Applying  $M_{A_1A_2}$ to $|\chi_3\rangle$ we result in
\begin{eqnarray}\label{chi4}
|\chi_4\rangle=M_{A_1A_2}|\chi_3\rangle| =
|\chi_{out}\rangle_S |0\rangle_R |1\rangle_{A_1} ,\;\;
|\chi_{out}\rangle_S=
C^{-1} \sum_{i=0}^{N^{(S)}-1} \sum_{j=0}^{N^{(R)}-1} a_{ij}|i\rangle_S,
\end{eqnarray}
where $C$ provides the normalization, i.e.,
$C=\sqrt{\sum_{i=0}^{N^{(S)}-1}\left(\sum_{j=0}^{N^{(R)}-1}\right)^2}$. This algorithm  allows us to avoid the problem of small probability of required outcome after the ancilla measurement.
The depth of the described algorithm is determined by the operator $W^{(1)}_{RA_1}$ and equals $O(n^{(R)})$.
The  circuit realizing the described algorithm is given in Fig.\ref{Fig:CM}.

We shall note that, in general case of applying the controlled measurement  $M_{A_1A_2}$ to some superposition state $|\tilde \chi_3\rangle$ of a quantum system  with ancillae $A_1$ and $A_2$, the measurement operator $M_{A_2}$ is applied if only  $|\tilde \chi_3\rangle$ has at least one term with the ancilla state $|1\rangle_{A_1}$. Otherwise the state $|\tilde \chi_3\rangle $ remains unchanged: $M_{A_1A_2}|\tilde \chi_3\rangle =|\tilde \chi_3\rangle$.

We emphasize that the state  $|\chi_2\rangle$  in form (\ref{chi2}) appears in the algorithms of  inner product, matrix addition and multiplication,  determinant calculation, matrix inversion and linear system solver developed in Ref. \cite{ZQKW_arxiv2023} and  Ref.\cite{ZBQKW_2024}. We also have to note that the controlled measurement loses information related to the probability of obtaining the required state of the ancilla. In other words, the normalization $C$ in the state $|\chi_{out}\rangle_S$ remains undefined. However, this information is either not required or represents only a part of  result  in the set of algorithms  considered below.

\subsection{Application of controlled measurement to known algorithms}

We use the results of Ref. \cite{ZQKW_arxiv2023} and  Ref.\cite{ZBQKW_2024} in, respectively,  subsections \ref{Section:inpr}-\ref{Section:matrmult} and \ref{Section:det}-\ref{Section:lin}. In all those algorithm, there is already the one-qubit ancilla that labels the garbage. Thus, we always start with the state $|\chi_2\rangle$ in form (\ref{chi2}), add one more  one-qubit ancilla, then  use the C-NOT  defined in (\ref{cnot}) and finally apply the controlled measurement (\ref{CM}) ending up with the state $|\chi_4\rangle$ in  form (\ref{chi4}).

\subsubsection{Inner product}
\label{Section:inpr}
From Ref.\cite{ZQKW_arxiv2023} we have the state
\begin{eqnarray}
|\chi_2\rangle \equiv |\Phi_4\rangle  =\frac{\langle \Psi_2^*|\Psi_1\rangle}{2^{3 n/2}}
  |0\rangle_{S_1} |0\rangle_{S_2} |0\rangle_{A} |1\rangle_{B_1} |1\rangle_{B_2} +
	|g_2\rangle_{S_1S_2AB_1} |0\rangle_{B_2},
\end{eqnarray}
where we use the one-qubit  ancilla $B_2$ instead of $A_1$ to label the garbage.
We add one more one-qubit ancilla $A_2$ in the ground state
and apply the  C-NOT $W^{(2)}_{B_2A_2}$ (\ref{cnot})
to the state $|\chi_2\rangle |0\rangle_{A_2}$:
\begin{eqnarray}
|\chi_3\rangle  =
W^{(2)}_{B_2 A_2}|\chi_2\rangle  |0\rangle_{A_2}  =\frac{\langle \Psi_2^*|\Psi_1\rangle}{2^{3 n/2}}
  |0\rangle_{S_1} |0\rangle_{S_2} |0\rangle_{A} |1\rangle_{B_1} |1\rangle_{B_2} |1\rangle_{A_2} +
	|g_2\rangle_{S_1S_2AB_1} |0\rangle_{B_2} |0\rangle_{A_2}.
\end{eqnarray}
Applying the controlled measurement $M_{B_2A_2}$ (\ref{CM})  to  $|\chi_3\rangle$
we obtain
\begin{eqnarray}
|\chi_4\rangle =M_{B_2A_2} |\chi_3\rangle  = \frac{\langle \Psi_2^*| \Psi_1\rangle}{|\langle \Psi_2^*| \Psi_1\rangle|  }  |0\rangle_{S_1} |0\rangle_{S_2} |0\rangle_{A}|1\rangle_{B_1}|1\rangle_{B_2},
\end{eqnarray}
which stores the phase of the inner product $\langle \Psi_2^*| \Psi_1\rangle$.

\subsubsection{Matrix addition}
\label{Section:matradd}
From Ref.\cite{ZQKW_arxiv2023} we have the state
\begin{eqnarray}
|\chi_2\rangle\equiv |\Phi_4\rangle=
	\frac{s}{2}\sum_{j=0}^{N-1}  \sum_{l=0}^{M-1} \Big( (a^{(1)}_{jl} +
	a^{(2)}_{jl})  {|j\rangle_{R_1} |l\rangle_{C_1} |0\rangle_{D_1}
		|0\rangle_{R_2} |0\rangle_{C_2} |0\rangle_{D_2}}\Big) |1\rangle_{B_1}  |1\rangle_{B_2}+
	|g_2\rangle_{R_1C_1D_1R_2C_2D_2B_1} |0\rangle_{B_2}.
\end{eqnarray}
Again, we use the one-qubit ancilla $B_2$ instead of $A_1$ to label the garbage.
We add one more one-qubit  ancilla $A_2$ in the ground state
and apply the  C-NOT $W^{(2)}_{B_2 A_2}$ (\ref{cnot}) to $|\chi_2\rangle |0\rangle_{A_2}$. Thus we obtain
\begin{eqnarray}
|\chi_3\rangle &=& W^{(2)}_{B_2A_2}  |\chi_2\rangle |0\rangle_{A_2}=
	\frac{s}{2}\sum_{j=0}^{N-1}  \sum_{l=0}^{M-1} \Big( (a^{(1)}_{jl} +
	a^{(2)}_{jl})  {|j\rangle_{R_1} |l\rangle_{C_1} |0\rangle_{D_1}
		|0\rangle_{R_2} |0\rangle_{C_2} |0\rangle_{D_2}}\Big) |1\rangle_{B_1}
 |1\rangle_{B_2} |1\rangle_{A_2}+\\\nonumber
&&
	|g_2\rangle_{R_1C_1D_1R_2C_2D_2B_1} |0\rangle_{B_2} |0\rangle_{A_2}.
\end{eqnarray}
Applying the controlled measurement $M_{B_2A_2}$ (\ref{CM}) to  $|\chi_3\rangle$
we have
\begin{eqnarray}\nonumber
&&
|\chi_4\rangle=M_{B_2A_2} |\chi_3\rangle  =
|\Psi_{out}\rangle |0\rangle_{D_1}
 |0\rangle_{R_2} |0\rangle_{C_2} |0\rangle_{D_2}|1\rangle_{B_1}|1\rangle_{B_2},\\\nonumber
 &&
|\Psi_{out}\rangle =G^{{ -1}} \sum_{j=0}^{N-1}  \sum_{l=0}^{M-1}   (a^{(1)}_{jl}+a^{(2)}_{jl})   |j\rangle_{R_1} |l\rangle_{C_1},
\end{eqnarray}
where $G$ is the normalization constant, $G{=(\sum_{jl} |a^{(1)}_{jl}+a^{(2)}_{jl}|^2)^{1/2}}$.
The result of addition is stored in the registers $R_1$ and $C_1$.

\subsubsection{Matrix multiplication}
\label{Section:matrmult}
From Ref.\cite{ZQKW_arxiv2023} we have the state
\begin{eqnarray}
|\chi_2\rangle=|\Phi_4\rangle  =
{\frac{1}{2^{3 k/2}}}\left({\sum_{j_1=0}^{N-1}}\sum_{j=0}^{K-1}\sum_{l_1=0}^{M-1}
a^{(1)}_{j_1 j}a^{(2)}_{j l_1}  |j_1\rangle_{R_1} |0\rangle_{C_1} |0\rangle_{R_2} |l_1\rangle_{C_2}\right)
|0\rangle_A|1\rangle_{B_1} |1\rangle_{B_2}+r_3 |g_3\rangle_{R_1C_1R_2C_2AB_1} |0\rangle_{B_2}.
\end{eqnarray}
Again, we use the one-qubit ancilla $ B_2$ instead of $A_1$ to label the garbage.
We add one more one-qubit ancilla $A_2$ in the ground state
and apply the  C-NOT $W^{(2)}_{B_2A_2}$ (\ref{cnot}) to $|\chi_2\rangle
 |0\rangle_{A_2}$. We obtain
\begin{eqnarray}
|\chi_3\rangle  &=& W^{(2)}_{B_2\tilde B_2} |\chi_2\rangle|0\rangle_{A_2}  =
{\frac{1}{2^{3 k/2}}}\left({\sum_{j_1=0}^{N-1}}\sum_{j=0}^{K-1}\sum_{l_1=0}^{M-1}
a^{(1)}_{j_1 j}a^{(2)}_{j l_1}  |j_1\rangle_{R_1} |0\rangle_{C_1} |0\rangle_{R_2} |l_1\rangle_{C_2}\right)
|0\rangle_A|1\rangle_{B_1} |1\rangle_{B_2}|1\rangle_{A_2}+\\\nonumber
&&
r_3 |g_3\rangle_{R_1C_1R_2C_2AB_1} |0\rangle_{B_2}  |0\rangle_{A_2}.
\end{eqnarray}
Applying the controlled measurement $M_{B_2A_2}$ to  $|\chi_3\rangle$
we obtain
\begin{eqnarray}\nonumber
&&
|\chi_4\rangle =M_{B_2A_2}|\chi_3\rangle=
|\Psi_{out}\rangle\, |0\rangle_{C_1} |0\rangle_{R_2}  |0\rangle_A|1\rangle_{B_1}|1\rangle_{B_2}  ,
\\\nonumber
&&
|\Psi_{out}\rangle =G^{{-1}} {\sum_{j_1=0}^{N-1}}\sum_{j=0}^{K-1}\sum_{l_1=0}^{M-1}
a^{(1)}_{j_1 j}a^{(2)}_{j l_1}  |j_1\rangle_{R_1} |l_1\rangle_{C_2},
G=\left(\sum_{j_1,l_2}\sum_j \Big| a^{(1)}_{j_1 j}a^{(2)}_{j l_1}\Big|^2 \right)^{-1/2},
\end{eqnarray}
where the $G$ is the  normalization constant, $G{=(\sum_{j_1,l_2} |\sum_j a^{(1)}_{j_1 j}a^{(2)}_{j l_1}|^2 )^{1/2}}$.
The result of multiplication is stored in the registers  $R_1$ and $C_2$.

\subsubsection{Determinant calculation}
\label{Section:det}
We use the formula from \cite{ZBQKW_2024}.
\begin{eqnarray}
|\chi_2\rangle\equiv |\Phi_3^{(D)}\rangle= W^{(2)}_{SAB} |\Phi_2^{(D)}\rangle|0\rangle_B 
=
 \frac{\det(\tilde A)}{2^{\tilde N/2}}\;
 |0\rangle_{S}|0\rangle_{A}
|1\rangle_{B}+
 |g_3\rangle_{SA}|0\rangle_B .
\end{eqnarray}
Here we use the one-qubit  ancilla $B$ instead of $A_1$ to label the garbage.
We  introducing the additional one-qubit ancilla $A_2$ in the ground state and then apply C-NOT $W^{(2)}_{BA_2}$ defined in (\ref{cnot})
to  $|\chi_2\rangle |0\rangle_{A_2}$. We obtain
\begin{eqnarray}
|\chi_3\rangle= W^{(2)}_{BA_2} |\chi_2\rangle |0\rangle_{A_2} = 
\frac{\det(\tilde A)}{2^{\tilde N/2}}\;
 |0\rangle_{S}|0\rangle_{A}
|1\rangle_{B} |1\rangle_{A_2}+
 |g_3\rangle_{SA}|0\rangle_B|0\rangle_{A_2} .
\end{eqnarray}
Applying the controlled measurement $M_{BA_2}$ (\ref{CM})  to $|\chi_3\rangle$ we obtain
\begin{eqnarray}
&&|\chi_4\rangle= M_{BA_2} |\chi_3\rangle=  |\Psi^{(D)}_{out}\rangle |0\rangle_{A}|1\rangle_{B} , \;\;  |\Psi^{(D)}_{out}\rangle =   \arg(\det(\tilde A)) \;
 |0\rangle_{S} , 
\end{eqnarray}
The only disadvantage of applying the controlled measurement is that we lose the information about the $|\det(A)|$ that was obtained in \cite{ZBQKW_2024} in terms of the probability of obtaining the ancilla $B$ in the state 1 after measurement.

\subsubsection{Matrix inversion}
\label{Section:inv}
From \cite{ZBQKW_2024} we have the following state:
\begin{eqnarray}\label{Phi2}
 |\chi_2\rangle\equiv|\Phi_4\rangle
=\sum_{{i=1}}^{N-1} \sum_{{j=1}}^{N-1} a_{ji}
|0\rangle_S  |j\rangle_R |i\rangle_C| 0\rangle_A |1\rangle_{B}+|g\rangle_{SA}|0\rangle_B,
\end{eqnarray}
where
\begin{equation} \label{ajl}
a_{ji} = -\frac{q\det(A) A^{-1}_{ji}}{2^{(\tilde N+n)/2}}.
\end{equation}
Again, we use the one-qubit ancilla $B$ instead of $A_1$ to label the garbage.
Introducing the one-qubit  ancilla $A_2$ in the ground state and applying the operator $W^{(2)}_{BA_2}$ to $ |\chi_2\rangle |0\rangle_{A_2}$  we obtain
\begin{eqnarray}\label{Phi2}
&& |\chi_3\rangle =  W^{(2)}_{BA_2} |\chi_2\rangle |0\rangle_{A_2}   =\sum_{{i=1}}^{N-1} \sum_{{j=1}}^{N-1} a_{ji}
|0\rangle_S  |j\rangle_R |i\rangle_C| 0\rangle_A |1\rangle_{B}|1\rangle_{A_2}+|g\rangle_{SA}|0\rangle_B|0\rangle_{A_2}.
\end{eqnarray}
Applying the controlled measurement $M_{BA_2}$  (\ref{CM}) to $|\chi_3\rangle$ we have
\begin{eqnarray}
|\chi_4\rangle =M_{BA_2}|\chi_3\rangle=|\Psi_{out}\rangle
|0\rangle_S | 0\rangle_A|1\rangle_{B} ,
\end{eqnarray}
where
\begin{eqnarray}
 \label{Psiout}
|\Psi_{out}\rangle &=& G^{-1}\sum_{{i=1}}^{N-1} \sum_{{j=1}}^{N-1} A^{-1}_{ji}   |j\rangle_R |i\rangle_C =G^{-1}  |A^{-1}\rangle,
\end{eqnarray}
where $G$ is the  normalization constant, $G=\sqrt{\sum_{i,j=1}^{N-1}|A^{-1}_{ij}|^2 }$.
Notice that the state $|A^{-1}\rangle=\sum_{{i=1}}^{N-1} \sum_{{j=1}}^{N-1} A^{-1}_{ji}   |j\rangle_R |i\rangle_C$ is not normalized.

\subsubsection{Linear system solver}
\label{Section:lin}
From \cite{ZBQKW_2024} we have the following state:
\begin{eqnarray}
|\chi_2\rangle\equiv |\Phi^{(L)}_6\rangle =
{\frac{1}{2^{ n/2}}}\left({\sum_{j_1=1}^{N-1}}\sum_{j=1}^{N-1}
a_{j_1 j}b_{j }  |j_1\rangle_{R} |0\rangle_{C} |0\rangle_{b} \right)|0\rangle_{B}   |1\rangle_{\tilde B}+
 |g_2\rangle_{RCb} |1\rangle_{B}  |0\rangle_{\tilde B}.
\end{eqnarray}
We use the one-qubit ancila $\tilde B$ instead of $A_1$ to label the garbage.
Introducing the one-qubit ancilla $A_2$ in the ground state and applying the operator $W^{(2)}_{\tilde BA_2}$ (\ref{cnot}) to $ |\chi_2\rangle |0\rangle_{A_2}$  we obtain
\begin{eqnarray}
|\chi_3\rangle =W^{(2)}_{\tilde BA_2} |\chi_2\rangle  |0\rangle_{A_2}=
{\frac{1}{2^{ n/2}}}\left({\sum_{j_1=1}^{N-1}}\sum_{j=1}^{N-1}
a_{j_1 j}b_{j }  |j_1\rangle_{R} |0\rangle_{C} |0\rangle_{b} \right)|0\rangle_{B}
 |1\rangle_{\tilde B} |1\rangle_{A_2}+
 |g_2\rangle_{RCb} |1\rangle_{B}  |0\rangle_{\tilde B} |0\rangle_{A_2}.
\end{eqnarray}
Applying the controlled measurement $M_{\tilde BA_2}$ (\ref{CM}) to $|\chi_3\rangle$ we have
\begin{eqnarray}
 &&
|\chi_4\rangle =M_{\tilde BA_2}|\chi_3\rangle=
|\Psi^{(L)}_{out}\rangle\, |0\rangle_{C} |0\rangle_{b}  |0\rangle_B |1\rangle_{\tilde B} ,
\\\nonumber
&&
|\Psi^{(L)}_{out}\rangle =G^{{-1}} {\sum_{j_1=1}^{N-1}}\sum_{j=1}^{N-1}
a_{j_1 j}b_{j }  |j_1\rangle_{R} =G^{{-1}} |x\rangle_R ,
\end{eqnarray}
where $G{=(\sum_{l,j} | a_{l j}b_{j }|^2 )^{1/2}}$ is the normalization constant.  Here $|x\rangle_R=\sum_{j=1}^{N-1}
a_{j_1 j}b_{j }  |j_1\rangle_{R}$ is an unnormalized vector.

\section{Hermitian conjugate}
\label{Section:HC}
\vspace{1cm}

\begin{figure}[ht]
    \includegraphics[width=0.2\textwidth]{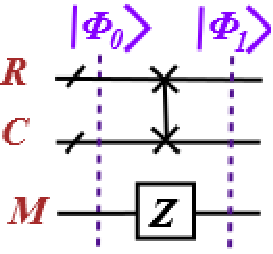}
    \caption{Circuit for Hermitian conjugation.}
\label{Fig:HC}
\end{figure}

We represent the complex elements of the $N\times N$ matrix $A=\{a_{ij}\}$ as
$a_{ij}=a_{ij0}+ i a_{ij1}$.
Let us encode the elements $a_{ij}$ into the superposition state of three subsystems.
Subsystems $R$ and $C$ encode the indexes of row and column respectively, each of them includes $\log N$ qubits. The third subsystem $M$ is a one-qubit subsystem. Its state is $|0\rangle$ for the real part of $a_{ij}$ and $|1\rangle$ for the imaginary part. Thus
the matrix $A$ is encoded into the state
\begin{eqnarray}
|\Phi_0\rangle=\sum_{m=0}^1 \sum_{i=0}^{N-1}\sum_{j=0}^{N-1}  a_{ijm} |i\rangle_R |j\rangle_C |m\rangle_M.
\end{eqnarray}
Then Hermitian conjugation can be realized by the following operator
\begin{eqnarray}
W=SWAP(R,C) Z_M,
\end{eqnarray}
where $SWAP$ exchanges states of the subsystems $R$ and $C$, and  $Z_M = \sigma^{(z)}_M$ is the $\sigma^{(z)}$-operator applied to $M$.
Thus
the matrix encoded into the state $W |\Phi_0\rangle$ is the Hermitian conjugate of $A$,
\begin{eqnarray}
|\Phi_1\rangle = W |\Phi_0\rangle = \sum_{m=0}^1 \sum_{i=0}^{N-1}\sum_{j=0}^{N-1}  a_{ijm} (-1)^m |j\rangle_R |i\rangle_C |m\rangle_M.
\end{eqnarray}
We emphasize that working with rectangular matrix $A$ asumes just putting zeros instead of  appropriate elements in the square matrix.
The circuit for this operation is shown in Fig.\ref{Fig:HC}.

We shall notice that using the new encoding of the complex matrix with the superposition state of the triple system $R\cup C\cup M$ requires appropriate modifications for the algorithm in Refs.\cite{ZQKW_arxiv2023,ZBQKW_2024}.

\section{Conclusions}
\label{Section:conclusions}
We introduce the controlled measurement as a tool allowing to resolve the problem of small access probability to the desired state of the ancilla with the purpose to remove the garbage acquired during performing the particular  quantum algorithm. This significantly increases the efficiency of the algorithms developed in Refs. \cite{ZQKW_arxiv2023,ZBQKW_2024}. Although we lose the part of information related to the access probability to the desired state of ancilla, because this probability is not measured in the modified algorithms. We also have to note that the method of practical realization of the controlled measurement is not clear yet.

In addition, we propose the modified encoding of the complex matrix into the superposition state of the triple system $R\cup C \cup M$, where the  basis states of the  subsystems $R$ and $C$ encodes the row and column indexes of the matrix elements while the  subsystem  $M$ serves as a label of the imaginary parts of the complex matrix elements. Such encoding allows to calculate the Hermitian conjugate  of an arbitrary matrix. Recall that the encoding of the matrix  into the system $R\cup C$ was already used in the algorithms of Refs. \cite{ZQKW_arxiv2023,ZBQKW_2024}.

\,

\,

\,

{\bf Acknowledgments.} The work was performed as a part of a state task, State Registration No. 124013000760-0. This work was also supported by the National Natural Science Foundation of
China (Grants No. 12031004, No. 12271474 and No. 61877054).

\end{document}